\documentstyle [12pt,epsfig] {article}

\newcommand{\be}{\begin{equation}}
\newcommand{\ee}{\end{equation}}
\newcommand{\beqs}{\begin{eqnarray}}
\newcommand{\eeqs}{\end{eqnarray}}
\def\({\left (}
\def\){\right )}

\begin{document}

\begin{titlepage}

\begin{flushright}
\begin{tabular}{l}
ITP-SB-98-72   \\
ANL-HEP-PR-98-140 \\
hep-th/yymmddd  \\
\end{tabular}
\end{flushright}

\vspace{8mm}
\begin{center}
Degeneration of ALF $D_n$ Metrics
\vspace{10mm}

G. Chalmers \footnote{email: chalmers@sgi2.hep.anl.gov},
M. Ro\v{c}ek \footnote{email: rocek@insti.physics.sunysb.edu},
S. Wiles \footnote{email: swiles@insti.physics.sunysb.edu}

\vspace{4mm}

Argonne National Laboratory \\
High Energy Physics Division \\
9700 South Cass Avenue \\
Argonne, IL 60439-4815 \\
\vspace{4mm}

Institute for Theoretical Physics  \\
State University of New York       \\
Stony Brook, N. Y. 11794-3840  \\

\vspace{20mm}

{\bf Abstract}
\end{center}

Beginning with the Legendre transform construction of hyperk\"ahler
metrics, we analyze the ALF version of the $D_n$ metrics.  We
determine the constraint equation obtained from extremizing the
$w$ coordinate of the generating function $F(z,\bar{z},u,\bar{u},w)$
and study its behavior as we send two of the mass parameters of the
$D_n$ metric to zero.  We find that the constraint equation enforces
the limit that the metric becomes that of multi-Taub-NUT.

\vspace{35mm}

\end{titlepage}
\newpage
\setcounter{page}{1}
\pagestyle{plain}
\pagenumbering{arabic}
\renewcommand{\thefootnote}{\arabic{footnote}}
\setcounter{footnote}{0}

\section{Introduction}

The ALF versions of the $D_n$ metrics have been shown
\cite{sen},\cite{chalmers},\cite{cherkis_kapustin} to be the
appropriate metrics for the moduli space of several interesting and
related theories.  Of perhaps most interest currently are their
relationship to M-theory.  They can be found as the metrics of
M-theory pertaining to an Atiyah-Hitchin space
with $n$ Kaluza-Klein monopoles.  In the M-theory compactification to type IIA string theory, this corresponds to a vacuum with an orientifold 6-plane and $n$ parallel D6-branes, respectively.
A type IIB interpretation is given in \cite{cherkis_kapustin}.  They
may also be found as metrics of the moduli space of $d=2+1$ $N=4$
supersymmetric $SU(2)$ Yang-Mills theory, with $n$ fundamental
hypermultiplets.  These two apparently different spaces are related.  If one probes the background of the type IIA string theory
described above with a D2-brane parallel to the D6-branes, the resulting low-energy theory on the D2-brane is that of
$N=4$ SYM with $n$ hypermultiplets.  The theory has as moduli
the $3$ scalars of the vector multiplet in $2+1$ dimensions, as well
as the dual photon associated with the $U(1)$ of the spontaneously
broken $SU(2)$.  It also has as parameters the $3$ masses
associated with each hypermultiplet.  These masses correspond in the type IIa theory to the positions of each corresponding brane in
the transverse space.  These metrics also appear in other contexts, such
as spontaneously broken $N=4$ $U(N)$ SYM
theory \cite{chalmers_hanany},\cite{hanany_witten},\cite{diaconescu},
and in monopole dynamics problems\cite{atiyah_hitchin}.

It is known \cite{sen} that the appropriate metric on M-theory with
one compact direction and $n$ KK monopoles is a ``superposition'' of the
Atiyah-Hitchin metric and the metric of Euclidean multi-Taub-NUT space
\cite{hawking}.  In the type IIA language, if a brane is taken to
infinity, or in the SYM language if the mass of the  hypermultiplet
is taken to infinity, then it decouples from the  theory and the
metric becomes $D_{n-1}$ ALF metric.  It was shown using QFT arguments
\cite{seiberg_witten} that if one places two of the branes onto the
orientifold plane ({\it{i.e.}},lets two masses vanish), the metric
becomes purely that of multi-Taub-Nut. In this paper, we try to see
how this behavior can arise geometrically from  the constraint equation of
the Legendre transform construction  \cite{martin_ivan}.  The form of
the multi-Taub-NUT metrics we find have an additional $Z_2$ symmetry that
combines with the  cyclic $C_k$ group to produce the dihedral group of the
$D_k$ series (i.e. the metric is unchanged by $\vec{r}\rightarrow
-\vec{r}$).

We begin in Section 2 by reviewing the Legendre transform
construction for the general ${\cal{O}}(2)$ and ${\cal{O}}(4)$
construction.  In Section 3, we derive and discuss the constraint
equation for the Kahler potential of the $D_n$ metrics in the Legendre
transform construction. In Section 4, we examine the
behavior of this constraint under the limit and see how it enforces
the condition on the metric $g_{D_n} \rightarrow g_{mTN_{2n}}$.
Finally, in Section 5 we discuss open problems and related issues.

\section{The ${\cal{O}}(2)$ and ${\cal{O}}(4)$ Legendre
transform constructions}

In this section we briefly review the generalized Legendre tranform
construction \cite{martin_ulf}.  We first consider the ${\cal{O}}(2)$
case.  We begin by defining $\zeta$ to be the coordinate on the Riemann
sphere.  Conjugation on the sphere is defined by the antipodal
mapping $\zeta \rightarrow {-1/ \bar{\zeta}}$.

In the Legendre construction we define a second-order polynomial
$\eta(\zeta) = \sum_a^{n=2} w_a \zeta^a$, i.e. a section of a
${\cal{O}}(2)$ line bundle over the sphere.  It obeys the
reality condition $\overline{\eta({-1 / \bar{\zeta}})} = -\zeta^{-2}
\eta(\zeta)$, which implies that $w_{2-a} = -(-1)^a \bar{w}_a$. This
constrains its form to be $\eta(\zeta) = \bar{z} + x \zeta -
z \zeta^2$, where $z$ is a complex coordinate and $x$ is a real
coordinate.

We define a function $F$ as follows

\be F(z,\bar{z},v,\bar{v},x) = {1 \over {2 \pi i}} \oint_C {{d\zeta}
\over \zeta^2} G(\eta,\zeta)
\label{generating function}
\ee
\noindent
where $G(\eta(\zeta),\zeta)$ is some arbitrary function of
the two variables $\zeta$ and $\eta$ and $C$ is an appropriately
chosen contour.

The Kahler potential ${\cal{K}}(z,\bar{z},u,\bar{u})$
is constructed from $F$ by means of the Legendre transformation

\be {\cal{K}}(z,\bar{z},u,\bar{u}) = F(z,\bar{z},x) - (u + \bar{u}) x
\ee
\noindent
where $x(z,\bar{z},u,\bar{u})$ is a solution to the minimization

\be {{\partial F} \over {\partial x}} = u + \bar{u}  \ee
\noindent
The metric may then be constructed as usual from the Kahler potential
in terms of the coordinates $z,\bar{z},u,\bar{u}$ by the
appropriate second derivatives \cite{martin_ulf}.

As an example relevant to this paper, we construct of the
multi-Taub-NUT metrics.  In this case

\be F = -\oint_{C_0} {{d\zeta} \over {2 \pi i \zeta}} {\eta^2(\zeta)
\over \zeta^2} + \oint_{C^{'}} {{d\zeta} \over \zeta}
{\eta(\zeta)\ln(\zeta)\over \zeta} - \sum_{i=1}^{n} \sum_{+,-}
\oint_{C_i} {{d\zeta} \over {2 \pi i \zeta^2}} (\eta(\zeta) \pm
\chi_i(\zeta))
\ln(\eta(\zeta) \pm \chi_i(\zeta))
\label{multi-Taub-Nut}
\ee
\noindent
Here $\chi(\zeta)_i$ are ${\cal{O}}(2)$ polynomials with constant
coefficients.  The contour $C^{'}$ is a lemniscate enclosing the two
roots of $\eta(\zeta)$, while the $C_i$ are lemniscates enclosing the
roots of $\eta(\zeta) \pm \chi_i(\zeta)$.  The contour $C_0$ is a simple
contour enclosing the origin.  All $4k$-dimensional metrics admitting $k$
tri-holomorphic isometries may be  constructed in terms of $k$
independent $O(2)$ multiplets \cite{hklr}.

Other metrics may be constructed in terms of higher order
polynomials $O(2n)$ satisfying a reality constraint \cite{martin_ulf}.
In the ${\cal{O}}(4)$ case, we define a polynomial on the sphere
$\eta(\zeta) = \sum_a^{n=4} w_a \zeta^a$.  We require this polynomial
to obey the reality condition $\overline{\eta({-1 / \bar{\zeta}})} =
\zeta^{-4} \eta(\zeta)$, which implies $w_{4-a}=(-1)^a \bar{w}_a$.
These conditions allow us to write the polynomial as $\eta(\zeta) =
\bar{z} + \bar{v} \zeta + x \zeta^2 - v \zeta^3 + z \zeta^4$, where
$z$ and $v$ are complex and $x$ is real.

As before $(\ref{generating function})$, we define a function $F$ and
perform the following Legendre transform to obtain the Kahler potential

\be {\cal{K}}(z,\bar{z},u,\bar{u}) = F(z,\bar{z},v,\bar{v},x) - uv
-\bar{u}\bar{v} \ee
\noindent
where $v(z,\bar{z},u,\bar{u})$; in this case we must eliminate
the auxiliary coordinate $x(z,\bar{z},u,\bar{u})$ via the
solution to
\beqs
&{\displaystyle {{\partial F} \over {\partial v}} = u \; \; , \;
\; {{\partial F} \over {\partial \bar{v}}} = \bar{u} }&\\
&{\displaystyle {{\partial F} \over {\partial x}} = 0 }&
\label{constreq}\eeqs
\noindent
Solving this constraint equation is difficult in practice, as
generally one obtains transcendental equations (in the
coordinates chosen above).

Without solving for the constraint equation, the construction of
a particular metric on a given hyperk\"ahler manifold becomes the
question how to find the proper function $G$ and
contour $C$, {\it{i.e.}}, the function $F$.  The metrics we
wish to study were determined in \cite{cherkis_kapustin} using
twistor methods and studied previously in \cite{martin_ivan}
and \cite{chalmers} using twistor space methods.

In \cite{seiberg_witten}, it was shown that if two hypermultiplet masses
vanish in the $D_n$ metrics for $n>2$, then the metric becomes a
multi-Taub-Nut that describes the perturbative  correction to the
low-energy theory on the brane (the instanton contributions vanish).
Physically, this is relatively easy to understand in the M-theory
language as the cancelling of the Ramond charges of the two 6-branes
with that of the orientifold 6-plane as they approach each other,
flattening the Atiyah-Hitchin metric and leaving only the Taub-Nut metric
of the remaining branes.  In the Legendre construction language, how this
phenomenon can arise is much more mysterious.  The
$D_n$ metrics are constructed from the ${\cal{O}}(4)$ construction,
whereas the multi-Taub-Nut metrics are constructed from the
${\cal{O}}(2)$ construction, which have different structures (for
example, the latter has a triholomorphic isometry).  How
the vanishing of two of the masses in the constraint equation of the
${\cal{O}}(4)$ construction could transform the Kahler potential into
one derivable from an ${\cal{O}}(2)$ construction is not self-evident.
In this note, we will show how the limiting behavior of the constraint
equation results in precisely such behavior.

\section{The Constraint Equation}

The generating functional for the ALF-$D_n$ metrics was suggested and
supported originally in \cite{martin_ivan},\cite{chalmers} and was
later proven in \cite{cherkis_kapustin} to be

\be
F = -\oint_{C_0} {{d\zeta} \over {2 \pi i \zeta}} {\eta(\zeta)
\over \zeta^2} + \oint_{C} {{d\zeta} \over \zeta} { \sqrt{\eta(\zeta)}
\over \zeta} - \sum_{i=1}^{n} \sum_{+,-} \oint_{C_i} {{d\zeta} \over
{2 \pi i \zeta^2}} (\sqrt{\eta(\zeta)} \pm \chi_i(\zeta))
\ln(\sqrt{\eta(\zeta)} \pm \chi_i(\zeta)) \ .
\label{genfunc}
\ee
\noindent

The $\chi_i$ are ${\cal{O}}(2)$ polynomials in $\zeta$, parameterizing
the deformations in the $D_k$ ALF metric. The metric is unchanged
by reversing the sign of the deformation parameters $\chi_i
\rightarrow -\chi_i$.  They obey the
reality condition defined in the previous section for ${\cal{O}}(2)$
polynomials and can expressed as $\chi_{i}(\zeta) = \bar{p_{i}} +
q_{i} \zeta - p_{i} \zeta^2$, where $p_i$ are complex parameters and
$q_i$ are real parameters.  They parametrize the position of the
monopoles (branes) in the theory, or equivalently, the masses of the
hypermultiplet.

Because of the reality condition, the roots of $\eta(\zeta)$ always
come in pairs $\alpha, -1 / \bar{\alpha}$ and $\beta, -1 /
\bar{\beta}$.  The contour integrals are easier to work with if we
rewrite $\eta$ as follows

\be \eta(\zeta) = \rho (\zeta-\alpha) (\zeta - \beta) ( \bar{\alpha}
\zeta + 1 ) (\bar{\beta} \zeta +1) \ee
\noindent
where $\rho$ is real and independent of $\zeta$.  We similarly
rewrite each $\chi_i$ in terms of a real scale $\sigma_i$ and a
complex root $r_i$ as follows

\be \chi_{i} (\zeta) = \sigma_i (\zeta-r_i) (\bar{r_i} \zeta + 1) \ee

The contour $C_0$ is defined as a simple contour enclosing the $\zeta$
origin counterclockwise.  The contour $C$ is defined as two simple
contours around the branch cuts of $\sqrt{\eta(\zeta)}$ as shown below

\begin{center}
\epsfig{file=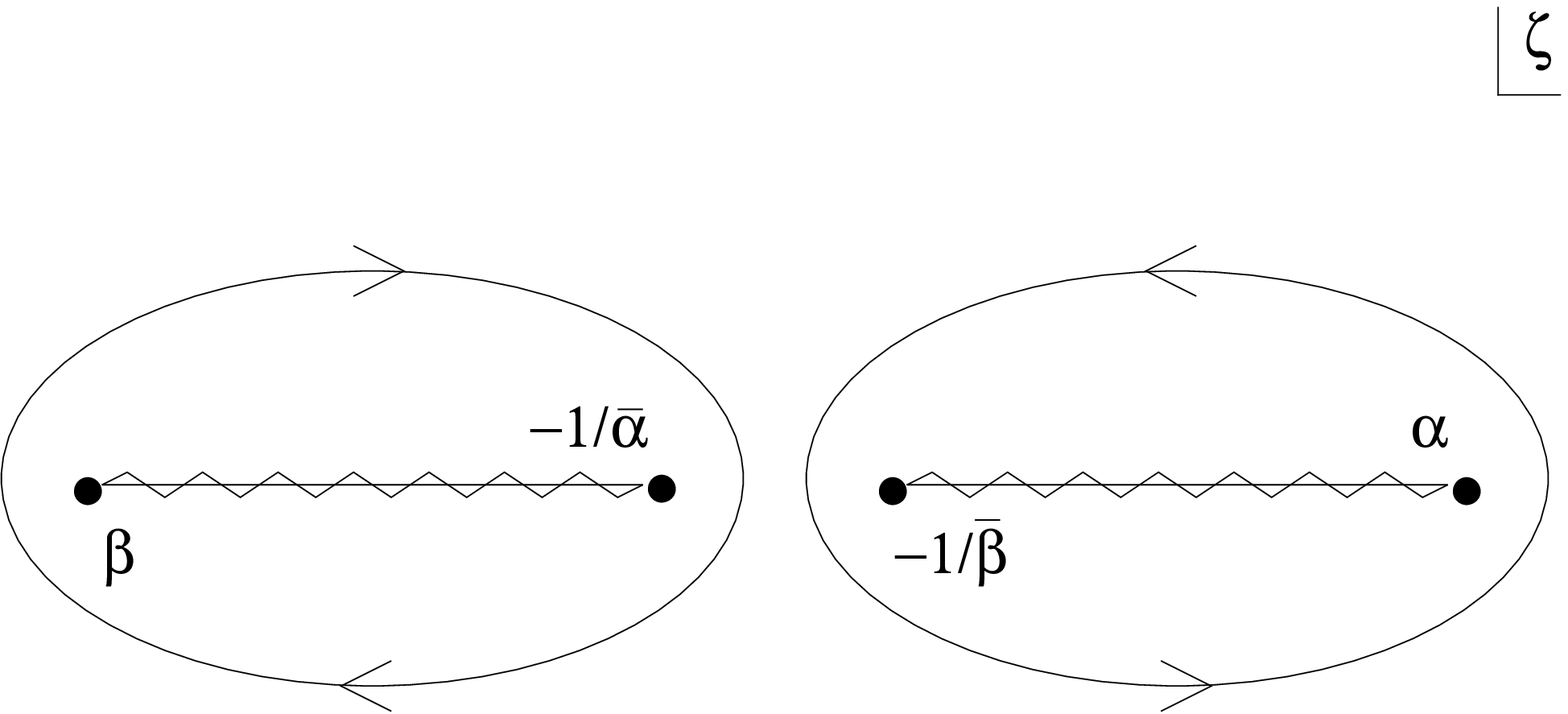,height=5cm,width=10cm}
\end{center}
\noindent
The contours $C_i$ are represented below

\begin{center}
\epsfig{file=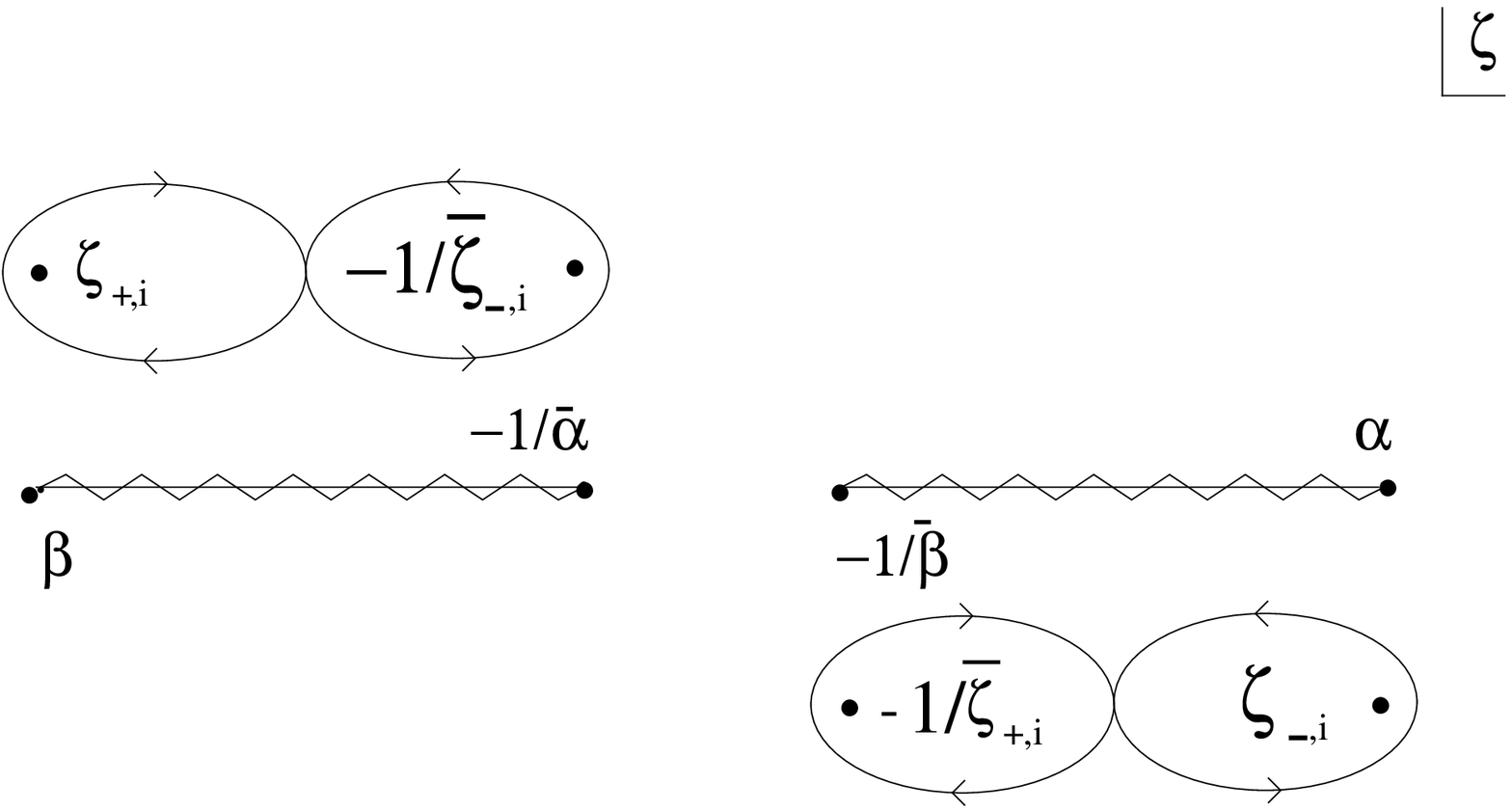,height=5cm,width=10cm}
\end{center}
\noindent
where the lemniscates are defined to encircle the roots of
$\sqrt{\eta} - \chi_i$ and $\sqrt{\eta} + \chi_i$, labelled above as
$\zeta_{+,i}, -1 / \bar{\zeta}_{-,i}$ and $\zeta_{-,i}, -1 /
\bar{\zeta}_{+,i}$ respectively.  The orientation
is chosen so that when these roots correspond to the roots of
$\eta(\zeta)$, they exactly cancel the contribution from $C$.

By the Legendre transform construction outlined previously, we must
extremize the generating function with respect to the coordinate $x$.
The resulting constraint equation is

\beqs \nonumber {\partial F} \over {\partial x} & = & 0 \\ & = &
-\oint_{C_0} {{d\zeta} \over {2 \pi i \zeta}} + {1 \over 2} \oint_{C}
{ {d\zeta} \over \sqrt{\eta(\zeta)} } - \sum_{i=1}^{n} \oint_{C_i} {
{d\zeta} \over \sqrt{\eta(\zeta)} } \ln( \eta(\zeta) - \chi_i(\zeta)^2
) \eeqs
\noindent
The first contour integral gives $-1$.  The other
contour integrals are expressible as a sum of complete and incomplete
elliptic integrals and yield the following constraint equation

\be 0 = -1 + { { (4-2n) K(k) + 2 \sum_{i=1}^n Re( F(\phi_i,k) +
F(\psi_i,k) ) } \over {\sqrt{\rho (1+\alpha\bar{\alpha})
(1+\beta\bar{\beta})} } } \ee
\noindent
where $K(k)$ is the complete elliptic integral of the first kind and
$F(x,k)$ is the incomplete elliptic integral of the first kind
\footnote{The definition of the incomplete elliptic integral of the
first kind $F(x,k)$ which we will use is, for real values of the
argument $x$ and the modulus $k$,

\be F(x,k) = \int_0^x {dt \over { \sqrt{1-t^2} \sqrt{1-k^2 t^2} }} \ee
\noindent
Since the arguments which appear in the constraint equation for this
function are in general complex, we take the analytic continuation of
$F(x,k)$ to complex values of $x$.  The modulus $k$ is always real in
our equations.  The complete elliptic integral is defined
as $K(k)=F(1,k)$.}.  The arguments of the incomplete elliptic
integrals are

\be \phi_i = \sqrt{ {(1+\beta \bar{\beta}) (a_i - \alpha)} \over {(1+
\beta \bar{\alpha}) (a_i - \beta)} } ~,~ \psi_i = \sqrt{ {(1+\alpha
\bar{\alpha}) (b_i - \beta)} \over {(1+\alpha\bar{\beta}) (b_i -
\alpha)} }
\label{ugh}
\ee
\noindent
and the modulus is

\be k = \sqrt { {(1+\alpha\bar{\beta})(1+\beta\bar{\alpha})} \over
{(1+\alpha\bar{\alpha})(1+\beta\bar{\beta})} } \ee
\noindent
Here $a_i$ and $b_i$ are respectively the two roots of the equation
$\eta - \chi_i^2 = 0$ analytically connected to the original roots
$\alpha$ and $\beta$, and are considered to be implicit functions of
$\alpha$,$\beta$, $\rho$, $\sigma_i$, and $r_i$.

As in \cite{martin_ivan}, it is most convenient to attempt to solve
for this equation in terms of the variable $\rho$ rather than the
original variable $x$, resulting in the final form of the constraint
equation we work with

\be \sqrt{\rho} = {{(4-2n) K(k) + 2 \sum_{i=1}^n Re( F(\phi_i,k) +
F(\psi_i,k) )} \over {\sqrt{(1+\alpha\bar{\alpha})
(1+\beta\bar{\beta})}}}
\label{constraint equation}
\ee
\noindent
Unfortunately, as the variables $a_i$ and $b_i$ are implicitly
functions of the variable $\rho$, we cannot obtain an analytic
solution for $\rho$:  The $\phi_i$ and $\psi_i$ are functions
of $\rho$ through equation (\ref{ugh}) and the solution for $\rho$ in
this set of coordinates is transcendental.

\section{Behavior of the Constraint Equation}

To begin our study of the behavior of the constraint equation, it is
useful to first consider its behavior when we take any given parameter
$\chi_i$ to be infinitely large.  Physically, this corresponds to
setting the masses of the $i^{th}$ hypermultiplet to infinity
\cite{seiberg_witten}, or to taking the transverse position of the
$i^{th}$ 6-brane to spatial infinity \cite{chalmers},\cite{sen}.  We
must therefore find that the resulting constraint equation is that of
a system with the $i^{th}$ monopole (6-brane) removed, {\it{i.e.}},
we must find that

\be Re( F(\phi_i,k) + F(\psi_i,k) ) \rightarrow 2 K(k)
\label{decoupling limit}
\ee
\noindent
where the left-hand side approaches $K(k)$ monotonically from below
\footnote{The limit must always be from below or there would be some
finite distance (mass) for which a brane (hypermultiplet) completely
decoupled from the theory, which is not physically reasonable.}.

For simplicity, consider taking the limit by sending $\sigma_i$ to
infinity.  It is clear that as $\sigma_i \rightarrow \infty$, the
roots of $\eta-\chi_i^2$ approach those of $\chi_i$.  For real roots
$\alpha$, $\beta$ and $r_i$, it is easy enough to use the addition
properties of incomplete elliptic integral to prove $(\ref{decoupling
limit})$, and the result analytically continues.  One may, of course,
continue to decouple all of the monopoles, which will leave the
constraint equation for the Atiyah-Hitchin space \cite{martin_ivan}.

The asymptotic regime
occurs when the ${\cal{O}}(4)$ coordinate degenerates into
${\cal{O}}(2)^2$.  The metric is a double cover of (the singular)
multi-Taub-Nut.  The  latter is described by eq.\ (\ref{multi-Taub-Nut}).
The metric may be easily derived.

The behavior of the $D_n$ metrics presented in
\cite{seiberg_witten},\cite{sen} is that for $n>2$, if one sets two
of the $\chi_i$ to zero, the metric of the moduli space should be the
multi-Taub-Nut metric \cite{sen}.  The asymptotic form of the
$D_n$ ALF metric is,
\be
ds^2 = V({\vec r},{\vec r}_i) d{\vec r} \cdot d{\vec r} +
V^{-1}({\vec r},{\vec r}_i) (d\phi + \vec A\cdot d\vec r)^2 \ ,
\ee
where $r=\sqrt{x^2+4 z \bar{z}}$
({\it cf}.\ eq.\ ($\ref{multi-Taub-Nut}$)),
\be
V({\vec r},{\vec r}_i)=1 - {4\over \vert {\vec r}\vert} +
\sum_{k=1}^n {1\over \vert {\vec r}-{\vec r}_i\vert} +
{1\over \vert {\vec r}+{\vec r}_i\vert} \ ,
\label{asymptotic}
\ee
and ${\vec\nabla}\times {\vec A} = {\vec \nabla} V$.  The form
of this metric is singular at some finite radius where $V=0$ (of course,
this singularity is not present in the full $D_k$ ALF metric, just in the
asysmptotic form).  However, when we take two of the
${\vec r}_i$  to zero, the contribution cancels the negative term in $V$
and forces the net result to be smooth.

In the language of the Legendre tranform construction, this
corresponds to the constraint equation enforcing the condition
$\eta_4(\zeta) \rightarrow \tilde{\eta}_2(\zeta)^2$ for {\it{fixed}}
$\rho$, where $\tilde{\eta}_2$ is an ${\cal{O}}(2)$ polynomial of $\zeta$.
Unfortunately, it is not possible to prove this by setting two
$\chi_i$ to zero identically.  If one does so, the corresponding roots
of $\eta - \chi_i^2 = 0$ are simply $a_i=\alpha$ and $b_i=\beta$ and
the corresponding incomplete elliptic integrals vanish.  The incomplete
elliptic integral terms for the remaining $\chi_i$, however, may never
exceed $(2n-4) K(k)$ in value combined, and so the right-hand side of the
constraint equation is always negative.   Thus, the only solution to the
constraint equation for any values of $\alpha$, $\beta$, and the $n-2$
remaining $\chi_i$ necessarily must be a negative value of $\sqrt{\rho}$.
Such a value seems unphysical for reasons we discuss later, and we ignore
such solutions \footnote{Each point in $(z,v,x)$ space uniquely defined a
given polynomial $\eta(\zeta)$.  However, a point in $(\rho,\alpha,\beta)$
space only uniquely defines $\eta$ up to the set of permutations of the
roots with appropriate rescalings of $\rho$.  These appear as discrete
isometries in the metric.  The permutation $\alpha \rightarrow {-1 /
\bar{\alpha}} , \rho \rightarrow {-\rho / {\alpha\bar{\alpha}}}$ allows
us to consider only positive values of $\rho$.  The metric also has a
$U(1)$ non-triholomorphic  isometry that implies
$F(\rho,\alpha,\beta,\sigma_i,r_i)=F(\rho, e^{i
\theta}\alpha, e^{i \theta} \beta, \sigma_i,  e^{i \theta} r_i)$, and
which we use to consider only real values of $\alpha$ in the rest of this
paper \cite{boyerfinley},\cite{chalmers}.}.
Note, however, that negative values of $\sqrt{\rho}$ give the same
four solutions as positive values in (\ref{roots}), up to
permutations.   We must take more care and instead consider a limiting
procedure for taking two $\chi_i$ to zero.

Since the $a_i$ and $b_i$ are implicit functions of $\sqrt{\rho}$, how
can we best analyze this limit?  Consider the following.  The
variables $a_i$ and $b_i$ are by definition roots of the equation

\be
\rho(\zeta-\alpha)(\zeta-\beta)(\bar{\alpha}\zeta+1)(\bar{\beta}\zeta+1)
- \sigma_i^2(\zeta-r_i)^2(\bar{r_i}\zeta+1)^2 = 0 \ee
\noindent
which in turn means they are also roots of the equation \be
(\zeta-\alpha)(\zeta-\beta)(\bar{\alpha}\zeta+1)(\bar{\beta}\zeta+1) -
({\sigma_i \over \sqrt{\rho}})^2(\zeta-r_i)^2(\bar{r_i}\zeta+1)^2 = 0
\label{roots}
\ee
\noindent
It is easiest to study the behavior of the constraint
equation by scaling out $\rho$.  We use the following procedure:
Assume $\alpha$ and $\beta$ are fixed.  Fix all ratios of the
$\sigma_i$ with respect to each other and vary all $\chi_i$ by an
overall scale $S$. To this end, we define a set of ``scaled''
mass parameters $\tilde{\chi}_i$ as follows

\be \tilde{\chi}_i = \tilde{\sigma_i}(\zeta-r_i)(\bar{r_i}\zeta+1) \;
\; i=1 \ldots n \ee
\noindent
where ${\tilde{\sigma}_i \over \tilde{\sigma}_j}={\sigma_i \over
\sigma_j}$ for all $i,j=1..n$ and the smallest scale is equal to one.
We now find roots $a_i,b_i$ for the equation
$\eta|_{\rho=1}-\tilde{\chi}_i^2=0$.  In this approach, the right-hand
side of the constraint equation no longer implicitly depends on
$\rho$.  Once we have found a solution of $\sqrt{\rho}$ for a given
$S$, we may ``rescale'' $\tilde{\sigma}_i$ by and so find a consistent
solution of the constraint for fixed values of $\alpha,\beta$ and
parameters $\chi_i =\sqrt{\rho} S \tilde{\chi_i}$.  Finding a solution
for particular values of $\chi_i$ then becomes a matter of numerical
iteration.

Consider the cases when $n>2$.  When the scale $S$ is set to zero, all
of the incomplete elliptic integral are also zero, and the solution
to the constraint equation is $\sqrt{\rho} = {{(4-2n) K(k)} \over
{\sqrt{(1+\alpha\bar{\alpha})(1+\beta\bar{\beta})}}}$, a negative
number. As $S$ approaches infinity, all of the monopoles will steadily
decouple and the solution to the constraint equation approaches that
of the Atiyah-Hitchin metric, $\sqrt{\rho} \rightarrow {{4 K(k)} \over
{\sqrt{(1+\alpha\bar{\alpha})(1+\beta\bar{\beta})}}}$, a positive
number.  As the right-hand side of the constraint equation is a
continuous function of its arguments, there is some $S=S_0$ for which
the solution is $\sqrt{\rho}=0$.  For larger values of $S$, the value
of $\sqrt{\rho}$ is positive and steadily increases to the fully
decoupled value.  Thus, we find that we have consistent solutions for
$\chi=0$ at more than one point, namely $S=0$ and $S=S_0$.  The
existence of such double solutions holds for all but one of the
non-positive values of $\sqrt{\rho}$.  The following graph represents all
consistent solutions of the constraint equation $(\ref{constraint
equation})$ obtainable by scaling all three mass parameters $\chi_i$ with
fixed ratios for an $n=3$ case.  This graph is representative of the
dependence of $\sqrt{\rho}$ on the true scale of an arbitrary mass
parameter $\chi$ for all choices of $\alpha,\beta,$ and $\chi_i$, and we
did not include specific values of $\chi_i$ for simplicity.  Having found
multiple solutions of $\rho$ for the same mass parameters $\chi_i$ in the
$\sqrt{\rho} \leq 0$ region as well as an apparent maximum scale for the
mass parameters past which no solutions exist, we do not consider this
region to be physically meaningful.  For completeness, we have described
these ``extra'' solutions, but we do not believe they have any physical
significance.

\begin{center}
\epsfig{file=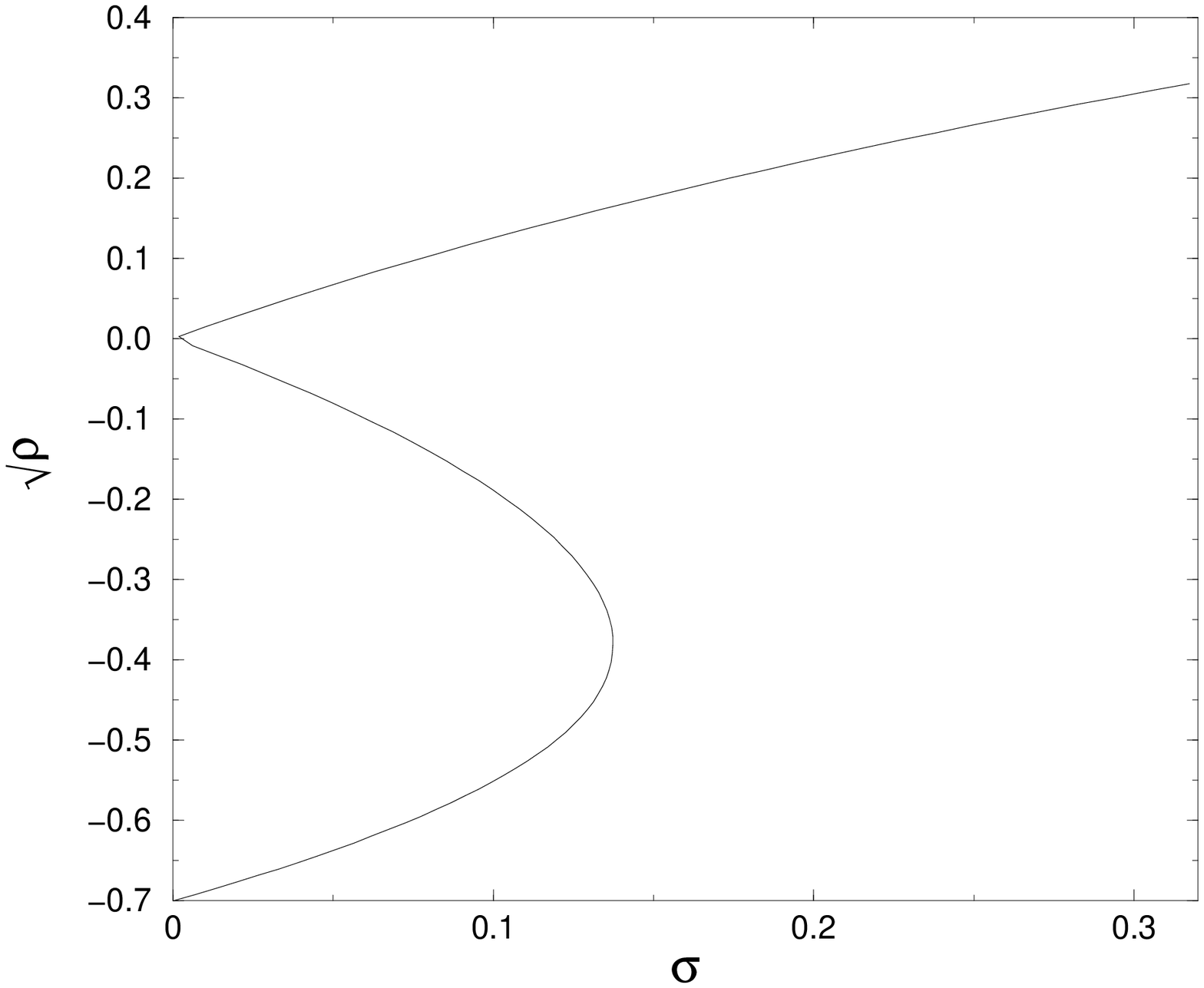,height=10cm,width=10cm}
\end{center}

To analyze the limit of taking two $\chi_i$ to zero for $\sqrt{\rho}$
positive while holding $\sqrt{\rho}$ and the other $\chi_i$ fixed,
consider the following procedure.  Hold $\alpha$, $\beta$, and all
ratios ${\sigma_i \over \sigma_j}, i,j=3 \ldots n$ fixed.   Define

\be \tilde{\chi}_i =  s (\zeta-r_i)(\bar{r_i}\zeta+1) \; \; i=1,2 \ee
\noindent
\noindent
where $s \ll 0$ is assumed to be a small real parameter we take to zero.
It is then always possible by our previous arguments to find
a particular value of $S > S_0$ such that $\sqrt{\rho} S
\tilde{\sigma_i} = \sigma_i, \; i=3 \ldots n$.  By varying the value
of $s$ incrementally to zero (without reaching zero, of course) one
may plot the behavior of $\sqrt{\rho}$ as a function of $s$ for fixed
values of $\alpha$, $\beta$.  We plotted several families of graphs
for $n=3$ with various fixed values of $r_i, i=1 \ldots n$ and
$\sigma_j,j=3 \ldots n$ and with values of $\beta$ steadily
approaching a fixed real $\alpha$ in each family.  Three such
>representative families of graphs are shown below.

In the following two families of graphs, the value of $\beta$ is
varied along the circle of radius $|\alpha|$ from graph to graph with
$\alpha$ fixed.  The legend indicates $|\alpha-\beta|$ for each graph,
as well as the roots and (relative) scales of the $\chi_i$.  One value
of $\beta$ widely separated from $\alpha$ as well as three with
$\beta$ closely approaching $\alpha$ are shown for each family.

\begin{center}
\epsfig{file=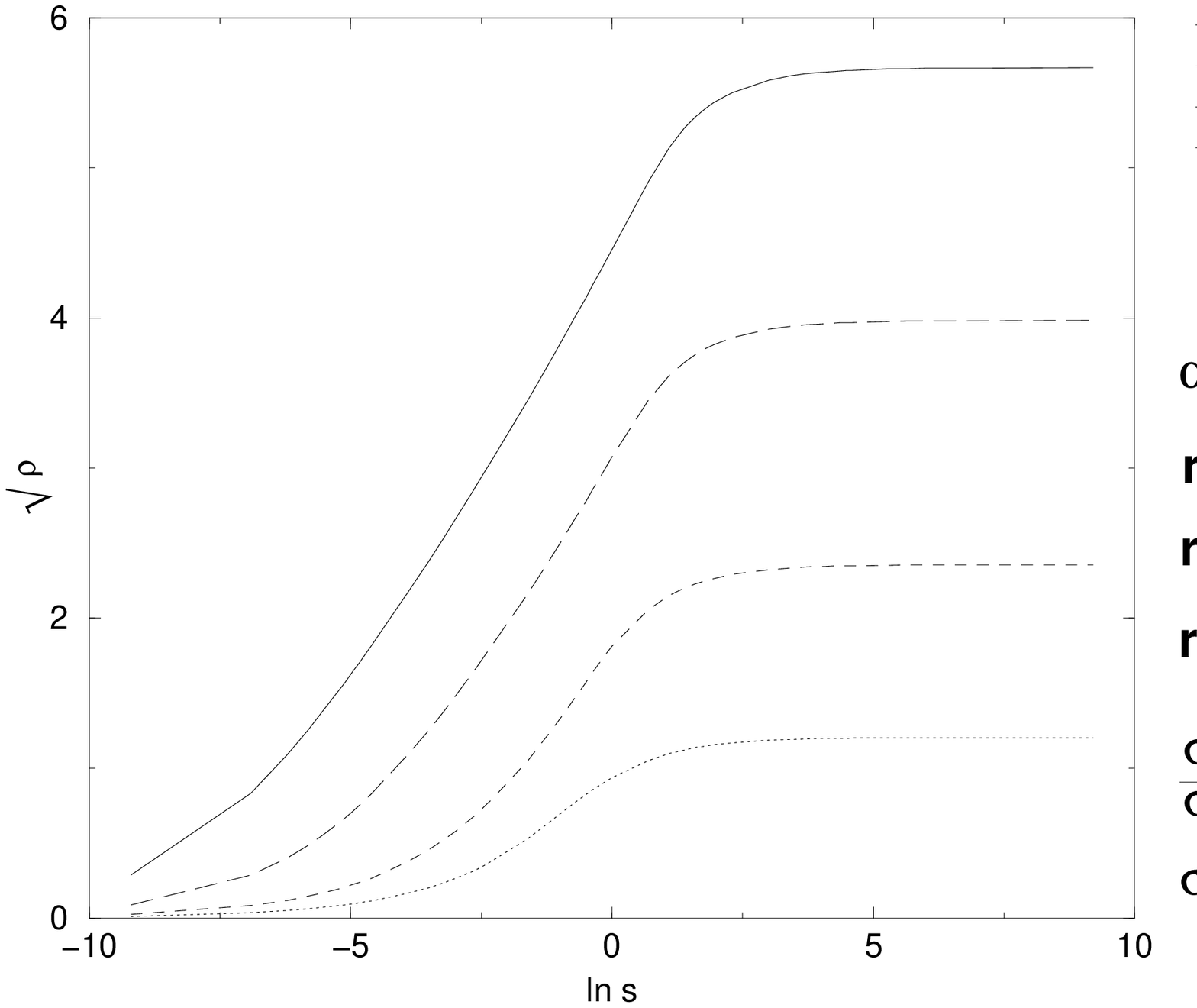,height=10cm,width=10cm}
\end{center}
\noindent

\begin{center}
\epsfig{file=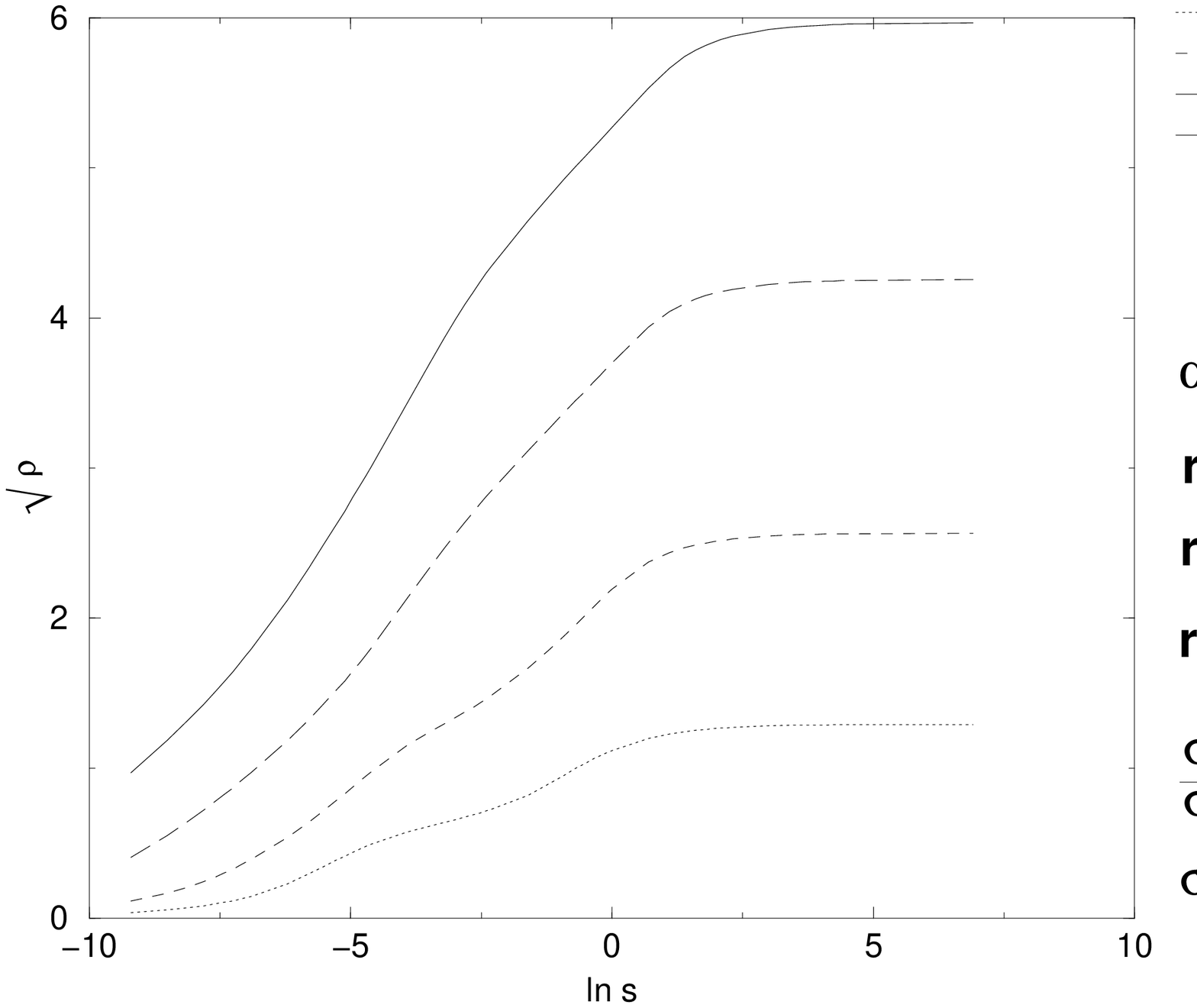,height=10cm,width=10cm}
\end{center}

In this graph, $\beta$ approaches $\alpha$ along the real axis
(from below).

\begin{center}
\epsfig{file=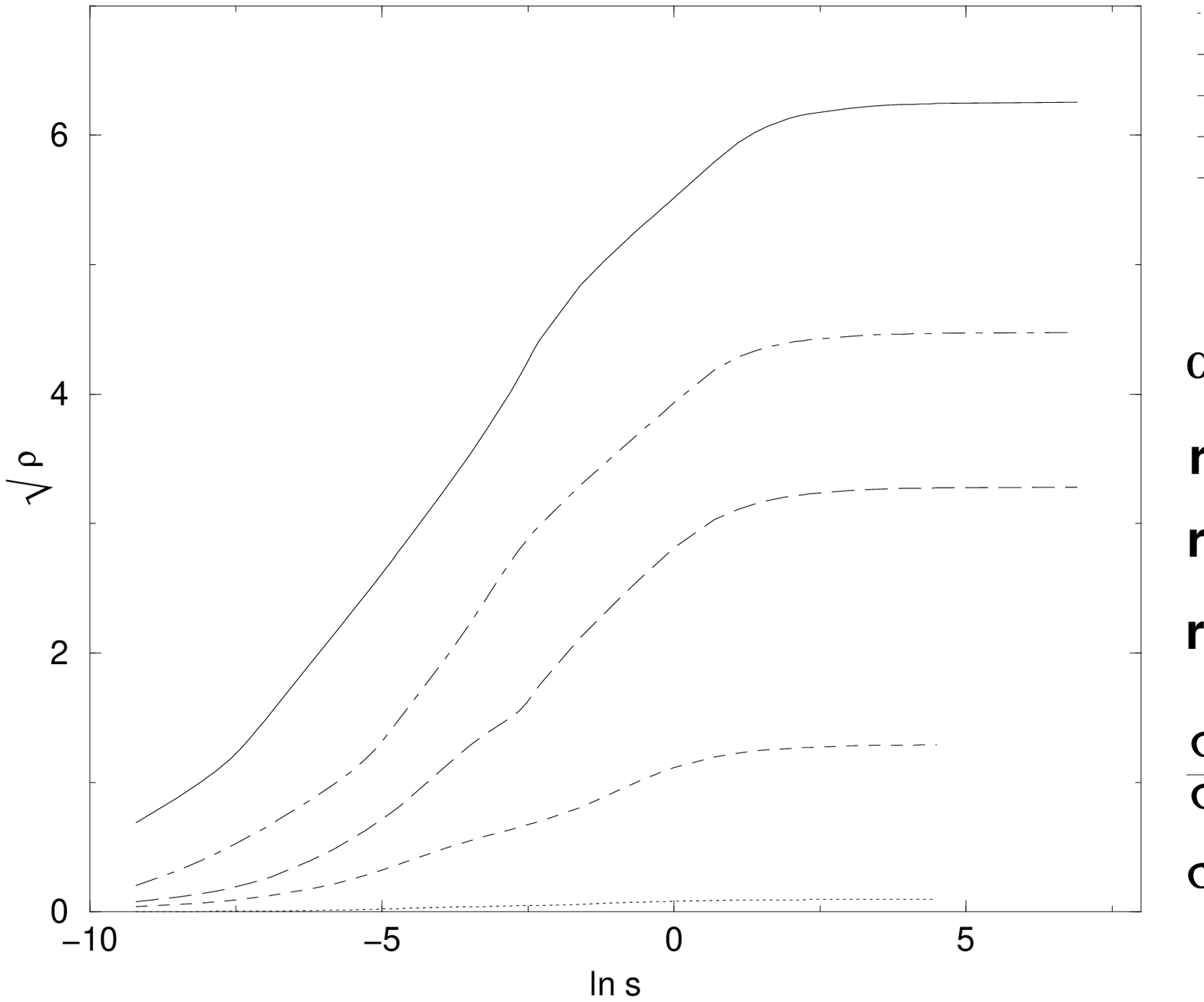,height=10cm,width=10cm}
\end{center}

\noindent
It is easy to see from these graphs that as $s \rightarrow 0$, one
must take values of $\beta$ ever closer to $\alpha$ to find a
solution for a given value of $\sqrt{\rho}$.  Of the other twenty-four
graphs which we plotted for various $\chi_i$, each displays the same
qualitative behavior.

Studying the behavior of the graphs as $s \rightarrow 0$ for fixed
$\sqrt{\rho}$, one finds that the absolute difference between $\alpha$
and $\beta$ must be linearly related to the scale $s$ to lowest order
when $s$ and $\epsilon=|\alpha-\beta|$ are small.  Given below is an
example graph of this relationship for the case where
$r_1=1+i,r_2=2+3i,r_3=1-2i$,$\sigma_3=4$, and $\alpha=2$

\begin{center}
\epsfig{file=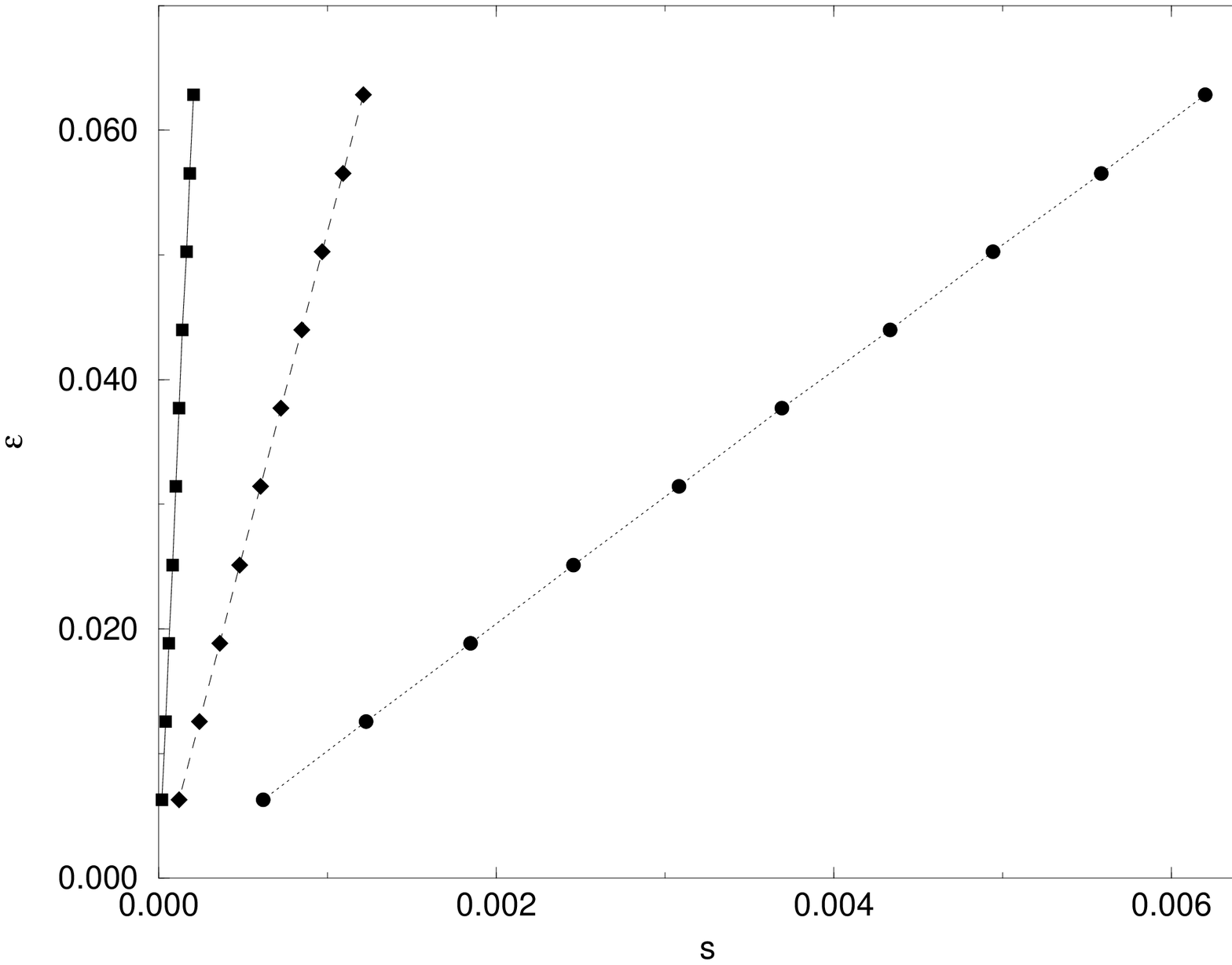,height=10cm,width=10cm}
\end{center}

This graph makes it clear that the constraint equation's intrinsic
behavior enforces the condition that as two mass parameters approach
zero, then $|\beta-\alpha| \rightarrow 0$ for a fixed $\rho$, and hence
$\eta_4\rightarrow \tilde{\eta}_2^2$.  (Taking more than two masses to
zero will also force the coincidence of $\alpha$ and $\beta$
for fixed $\rho$.)

We end this section with some implications of the above
degeneration.  The asymptotic form of the metric in (\ref{genfunc})
we are considering has the explicit form in (\ref{asymptotic})
in the normalization of the metric chosen in eq.\ (7).  This
metric has the form of the one-loop correction to the
vacuum moduli space metric of an $SU(2)$ $N=4$ gauge theory
gauge theory with $n_k$ hypermultiplets.  The form verifies
the predictions of \cite{seiberg_witten} that the instanton
contributions vanish when we take two of the mass parameters
to zero.

The novel feature of the suppression of instanton corrections as two masses tend to zero is the following:  the
instanton corrections are {\it not} explicitly suppressed in the
function $F$ or in $F_x$, but only in the {\it solution} to the
constraint equation $F_x=0$.

\section{Conclusions and Discussion}

We have shown how the $D_k$ ALF metrics become the Taub-NUT metric as
two masses tend to zero in the framework of the twistor construction of
\cite{martin_ivan, cherkis_kapustin}. The suppression of the
corrections that distinguish these metrics occurs in an interesting way:
in the limit as the masses tend to zero, the solution to the constraint
equation that arises in the construction of the $D_k$ metrics gets
``squeezed'' into the asymptotic domain where the corrections become
exponentially small.

Higher dimensional hyperk\"ahler ALF spaces analogous to the $D_k$ spaces
clearly limit in the same way to higher dimensional Taub-NUT spaces.
It would be interesting to analyze the constraint equation
(\ref{constreq}) for other cases, {\it e.g.,} $D_k$ ALE spaces.

Both the $A_k$ and the $D_k$ spaces are known to have descriptions as
algebraic curves \cite{kron}. It would also be interesting to see how the
limit arises in this description.

\vskip .3in
{\noindent\bf\large Acknowledgements}
\vskip .3in

We would like to thank S.\ Cherkis for discussions.
G.C. would like to thank CERN for its hospitality during the final
stages of this work.  The work of G.C. was supported in part by NSF
Grant No. PHY 9722101 and US Dept. of Energy, Division of High
Energy Physics, Contract W-31-109-ENG-38.  The work of S.W. and
M.R. was supported in part by NSF Grant No. PHY 9722101.

\vfill\break

\end{document}